\makeatletter
\@ifundefined{@parse@version@dash}{%
\def\@parse@version#1{\@parse@version@0#1}
\def\@parse@version@#1/#2/#3#4#5\@nil{%
\@parse@version@dash#1-#2-#3#4\@nil}
\def\@parse@version@dash#1-#2-#3#4#5\@nil{%
  \if\relax#2\relax\else#1\fi#2#3#4 }
}{}
\makeatother

\documentclass[10pt,aps,twocolumn,floats,floatfix,superscriptaddress,nofootinbib]{revtex4-2}
\usepackage[utf8]{inputenc}               		

\usepackage{graphicx,mathtools,amssymb,amsmath,amsthm,amsfonts,epsfig,epsf}
\usepackage[outdir=./]{epstopdf}
\usepackage[dvipsnames]{xcolor}
\usepackage[linktocpage]{hyperref}
\usepackage[mathscr]{euscript}
\usepackage{braket}
\usepackage{ragged2e}
\usepackage{subcaption}
\hypersetup{
    colorlinks=true,
    linkcolor=blue,
    filecolor=magenta,
    urlcolor=BrickRed,
    citecolor=BrickRed,
}
\hypersetup{pdfstartview=}
\usepackage{mathrsfs}

\usepackage{amsmath}

\newcommand{\be}{\begin{equation}}
\newcommand{\ee}{\end{equation}}
\newcommand{\beq}{\begin{eqnarray}}
\newcommand{\eeq}{\end{eqnarray}}

\begin{document}

\allowdisplaybreaks

\title{Quantum Gravity Induced Entanglement of Masses With Extra Dimensions. }


\author{Shuai Feng}
\affiliation{Department of physics, Nanchang University,
Nanchang 330031, China}
\affiliation{Center for Relativistic Astrophysics and High Energy Physics, Nanchang University, Nanchang, 330031, China}

\author{Bao-Min Gu}
\email{gubm@ncu.edu.cn}
\affiliation{Department of physics, Nanchang University,
Nanchang 330031, China}
\affiliation{Center for Relativistic Astrophysics and High Energy Physics, Nanchang University, Nanchang, 330031, China}

\author{Fu-Wen Shu}
\email{shufuwen@ncu.edu.cn}
\affiliation{Department of physics, Nanchang University,
Nanchang 330031, China}
\affiliation{Center for Relativistic Astrophysics and High Energy Physics, Nanchang University, Nanchang,
330031, China}
\affiliation{Center for Gravitation and Cosmology, Yangzhou University, Yangzhou, China}
%



\begin{abstract}

It is believed that gravity can be considered as a quantum coherent mediator. In this study, we propose a plan to test the existence of extra dimensions using the Quantum Gravity Induced Entanglement of Masses (QGEM) experiment. This experiment involves two freely falling test masses passing through a Stern-Gerlach-like device. We investigate the entanglement witness between these masses within the framework of the Randall-Sundrum II model (RS-II). Our findings indicate that the system reaches entanglement more rapidly in the presence of extra dimensions, particularly when the radius of the extra dimension is large.
\vfill
\end{abstract}


\maketitle

\newpage
\section{Introduction}


Extra dimensions present potential solutions to fundamental problems in theoretical physics, including the hierarchy problem \cite{arkani1998hierarchy,randall1999large}, the cosmological constant problem \cite{arkani2000small,dvali2003diluting,rubakov1983extra}, and the mass hierarchy of fermions \cite{kaplan2001new,yoshioka2000fermion}, among others. As a prominent type of theory beyond the standard model, extra dimension theories predict a wide range of phenomena in both high-energy and low-energy regimes. These novel physics phenomena are of great interest due to their potential to provide ways for detecting the existence of extra dimensions.


As a classical example, the Arkani-Hamed, Dimopoulos, and Dvali (ADD) model \cite{arkani1998hierarchy,arkani1999phenomenology,antoniadis1998new} offers an ingenious solution to the gauge hierarchy problem. In this model, the extra dimensions are flat and compactified in circles. The large radius of these dimensions leads to a modification of Newton's inverse square law at distances shorter than millimeters. The Kaluza-Klein modes of the graviton exhibit a mass gap of approximately $10^{-2}$ eV (for two extra dimensions), while the five dimension fundamental scale is the same as the electroweak scale. Consequently, the ADD model predicts numerous novel phenomena in high-energy particle physics and astrophysics processes.  However, this model has brought about a new hierarchy problem, that is, the problem between the magnitude of large extra dimension and the Planck length \cite{raychaudhuri2016particle}.

Another elegant scenario  to address this problem is the Randall-Sundrum type of theory \cite{randall1999large} (RS-I), where the extra dimension is curved and possesses an $S^1/Z^2$ symmetry. In this framework, gravitons, except for the massless mode, can be heavy and exhibit distinct phenomenologies in colliders compared to the large extra dimension theory. The corrections to Newton's law in this case depend on the $\mathrm{AdS}_5$ radius.  The phenomenological difference between the RS model and the ADD model mentioned above is due to the discrete spectrum produced by gravitons under the RS model \cite{raychaudhuri2016particle}. Similarly, S. Kumar Rai and S. Raychaudhuri also demonstrated the differences between the ADD model and the RS model in their work       \cite{rai2003single}.

Experiments aimed at detecting extra dimensions primarily rely on studying their phenomenologies. Current particle physics experiments have placed significant constraints on theories involving extra dimensions \cite{giudice1999quantum, battaglia2005contrasting}. However, constraints resulting from tests of Newton's inverse square law are relatively weaker \cite{spero1980test, adelberger2003tests, hoyle2001submillimeter, adelberger2001new, yang2012test, lee2020new}. In this study, we explore the possibility of detecting extra dimensions through entanglement experiments.

The operation in quantum teleportation is called one-way local operation and classical communication (LOCC), and it is very important in the quantum communication theory \cite{horodecki2009quantum}. That means if gravity is a classical operation, it cannot cause entanglement. This shows that if entanglement occurs between particles only under the action of gravity, then gravity should have certain quantum effects.

Recently, Bose et al. \cite{boseSpinEntanglementWitness2017} proposed a novel scheme to investigate entanglement between two freely falling masses in a gravitational field.  The experiment utilizes test mass with a very low internal crystal temperature and a device similar to Stern-Gerlach (SG) interferometers, which operate at extremely low temperatures to isolate the effects of factors other than gravity. The masses are initially released in a non-entangled state from a fixed height, and the entanglement witness is measured after a freely falling duration, denoted as $\tau$. This measurement captures gravitational information if the masses are in an entangled state, providing an opportunity to test corrections to Newton's inverse square law. Shielding non-gravitational effects presents a significant challenge in isolating the gravitational effect \cite{van2020quantum}, but such efforts are crucial for advancing our understanding of mass behavior, with implications spanning various fields in physics and beyond. Consequently, this area of research has garnered considerable attention in recent years \cite{elahi2023probing, li2023generation, matsumura2020gravity, schut2022improving, grossardt2020acceleration}.



Since the ADD model does not fully resolve the hierarchical problem, our work focuses on investigating the Quantum Gravity Induced Entanglement of Masses (QGEM) in the RS-II model \cite{randall1999alternative}. Our objective is to explore new approaches for studying extra dimensions. Our proposed strategy suggests that the presence of an extra dimension would result in the propagation of gravitons with $m>0$ along it. These correction terms would impact both the gravitational potential and particle entanglement. In this study, we propose incorporating the influence of extra dimensions into the QGEM experiment, allowing us to attempt to verify the existence of extra dimensions.

\section{RS Brane-Worlds}


There are two types of RS models: the RS-I model \cite{randall1999large} and the RS-II model \cite{randall1999alternative}. The RS-I model assumes that there are two 3-branes, one visible and the other invisible. Within this framework, the generation of the weak scale arises from  the order Planck scale through an exponential hierarchy. Notably, this exponential hierarchy is not a result of gauge interactions but is instead driven by the background metric, which corresponds to a slice of AdS$_{5}$ spacetime. The entire bulk is described by an action.
\begin{equation}
	\begin{aligned}
		&S = S_{gravity}+S_{vis}+S_{inv},\\	
		&S_{gravity}=\int d^{4}x\int dy\sqrt {-g}(2M_{5}^{3}R-\Lambda),\\
		&S_{vis}=\int d^{4}x \sqrt{-g_{vis}}(L_{vis}-V_{vis}),\\
		&S_{inv}=\int d^{4}x \sqrt{-g_{inv}}(L_{inv}-V_{inv}),
	\end{aligned}
\end{equation}
where
\begin{eqnarray}
	V_{vis}=-V_{inv}=-24M_{5}^{3}l^{-1},
\end{eqnarray}
and $M_{5}$ is the fundamental scale in five dimensions.
By using the action above and solving the 5D Einstein equation, one obtains
\begin{equation}
	\begin{aligned}
		&\sqrt{-G}\left ( R_{AB}-\frac{1}{2}G_{AB}R\right)=-\frac{1}{4M_{5}^{3}}\Big[\Lambda\sqrt{-G}G_{AB}\\ + &V_{vis}\sqrt{-g_{vis}}g_{\mu \nu }^{vis}\delta _{A}^{\mu }\delta _{B}^{\nu }\delta (\phi -\pi ) + V_{inv}\sqrt{-g_{inv}}g_{\mu \nu }^{inv}\delta _{A}^{\mu }\\
		&\delta _{B}^{\nu }\delta (\phi=0)\Big],
	\end{aligned}
\end{equation}
where $G_{AB}$ is the five dimensional metric and the indices $A,B=(\mu , \phi)$ $(\mu = 0,1,2,3)$. And $g_{\mu \nu }^{vis} \equiv G_{\mu \nu} (x^{\mu} , \phi = \pi  )$, $g_{\mu \nu }^{hid} \equiv G_{\mu \nu} (x^{\mu} , \phi = 0  )$.
The metric for the equation takes the form
\begin{equation}
	ds^{2}=e^{-2\left | y \right |/l}\eta _{\mu \nu }dx^{\mu }dx^{\nu }+dy^{2},
\end{equation}
where $\eta _{\mu \nu }$ has ``$-+++$'' signature. $y$ is the fifth compact dimension, and $dy=r_{c}d\phi$, with $r_{c}$ being the compactification radius of the extra dimension. 
 $e^{-2\left | y \right |/l}$ is called the warp factor and $l$ is the $AdS_{5}$ curvature radius of the extra dimension. This solution describes a curved extra dimension with a negative cosmological constant
\begin{equation}
	\Lambda=-24M_{5}^{3}l^{-2},\label{m5}
\end{equation}
which helps to confine the zero-mode gravitons near the invisible brane. The behavior of gravitons depends on their mass \cite{randall1999alternative,garriga2000gravity,giddings2000linearized,deruelle2001linearized}.

The RS-II model, on the other hand, assumes a 4+1 non-compactification dimensions and is to place one of the branes at $r_{c} = \infty $ \cite{randall1999alternative}. So the model derive
\begin{equation}
M_{Pl}=M_{5}^{3}l(1-e^{\frac{-2r_{c}\pi  }{l} } )=M_{5}^{3}l.
\end{equation}
The spectrum of general linearized fluctuations have the form
\begin{equation}
	G_{AB} =e^{-2\left | y \right |/l}\eta _{\mu \nu }+h_{\mu \nu } (x,y),
\end{equation}
and then do a variable decomposition $h(x,y)=\psi (y)e^{ip\cdot x} $. The gravitational fluctutations satisfy a wave equation
\begin{equation}
(\frac{-m^{2} }{2} e^{\frac{2|y|}{l}} -\frac{1}{2} \partial _{y}^{2} -2\frac{\delta (y)}{l} +\frac{1}{l^{2} })\psi (y)=0. \label{waveequation}
\end{equation}
Using the gauge $\partial ^{\mu }  h_{\mu \nu } =h_{\mu }^{\mu } =0$ and do a change of variables: $z\equiv sgn(y)\times (e^{\frac{|y|}{l}-1 } )l$, $\hat{\psi } (z) \equiv \psi(y)e^{\frac{|y|}{2l} }$, $\hat{h } (x,y) \equiv h(x,y)\times e^{\frac{|y|}{2l} }$. So the equation~(\ref{waveequation}) becomes
\begin{equation}
\left [ -\frac{1}{2} \partial _{z}^{2} +V(z)\right ]\hat{\psi }(z) =m^{2} \hat{\psi }(z) \label{waveequation2},
\end{equation}
where 
\begin{equation}
	V(z) = \frac{15}{8\left ( |z| + l \right )^{2} }-\frac{3}{2l}  \delta (z) ,
\end{equation}
and $\delta $ means the delta function. Under such conditions, 
the massless graviton recovers the Newton's inverse square law, while the massive modes produce a correction to the gravitational potential. The correction is
\cite {maartens2010brane,randall1999alternative}
\begin{equation}V\left ( r \right )\approx \frac{Gm_{1}m_{2}}{r}\left ( 1+\frac{2l^{2}}{3r^{2}} \right )\ (l < r),
	\label{correctGP}
\end{equation}

where $m_1$ and $m_2$ represent the masses of the test masses involved in the experiment

\section{The Effect of Extra Dimension}

\begin{figure}[h]
	\centering
	\includegraphics[width=8.5cm]{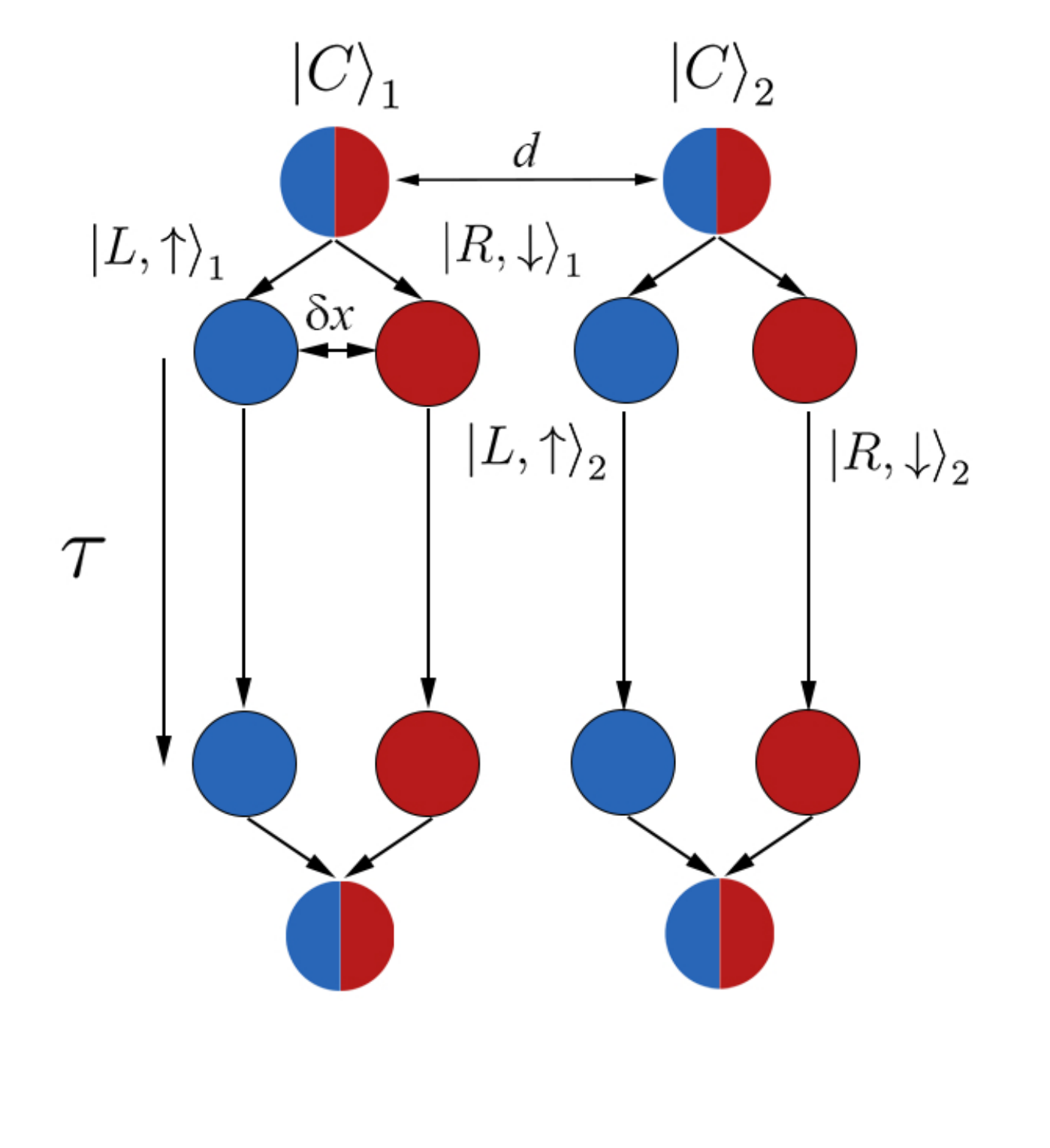}
	\captionsetup{justification=RaggedRight}
	\caption{The QGEM experiment postulates that two masses, separated by a distance $d$, undergo free fall for a duration of $\tau$ and become entangled solely through the gravitational force. Initially, the two test masses are in a superposition state characterized by localized states $\ket{L}$ and $\ket{R}$. They are then separated into $\ket{L,\uparrow}$ and $\ket{R,\downarrow}$ using an SG interferometer, and they remain in a superposition state during the period of free fall. Finally, their quantum states are measured after passing through a reverse SG interferometer. }
	\label{fig1}
\end{figure}

The QGEM experiment (see Fig.~\ref{fig1}) involves placing a pair of adjacent microobjects in a superposition state and allowing them to interact gravitationally for a duration $\tau$. It is expected that the system will be entangled in the final state \cite{boseSpinEntanglementWitness2017}. This entanglement can be confirmed by measuring the entanglement witness.


The QGEM experiment (see Fig.~\ref{fig1}) involves placing a pair of adjacent microobjects in a superposition state and allowing them to interact gravitationally for a duration $\tau$. It is expected that the system will be entangled in the final state \cite{boseSpinEntanglementWitness2017}. This entanglement can be confirmed by measuring the entanglement witness.

\begin{figure*}
	\centering
	\begin{subfigure}[t]{0.4\textwidth}
		\centering
		\includegraphics[width=7.0cm]{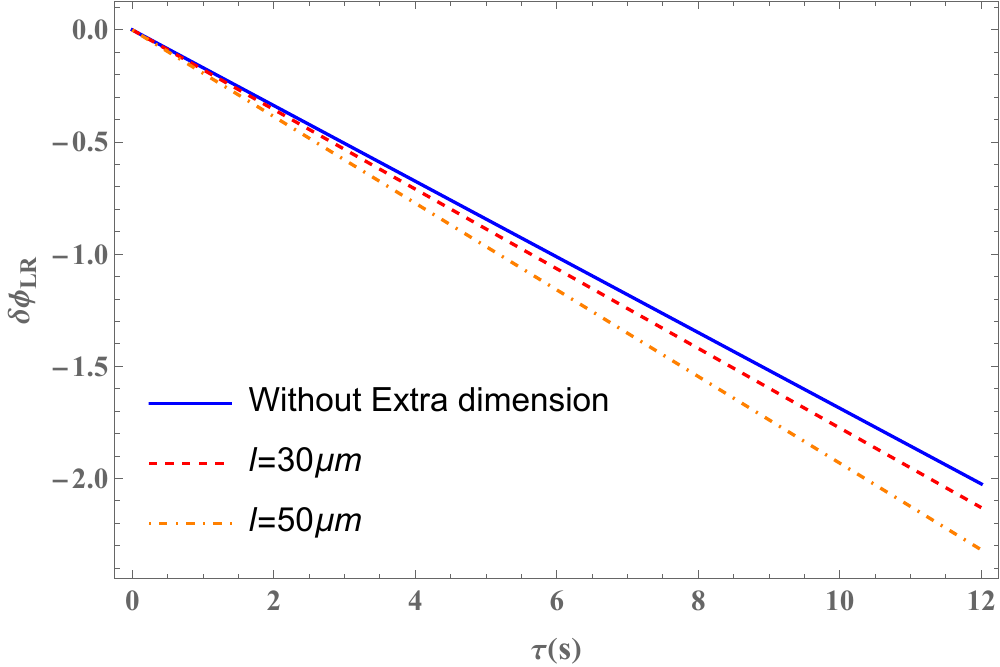}
		\caption{}
		\label{fig:a}
	\end{subfigure}
	\begin{subfigure}[t]{0.4\textwidth}
		\centering
		\includegraphics[width=7.0cm]{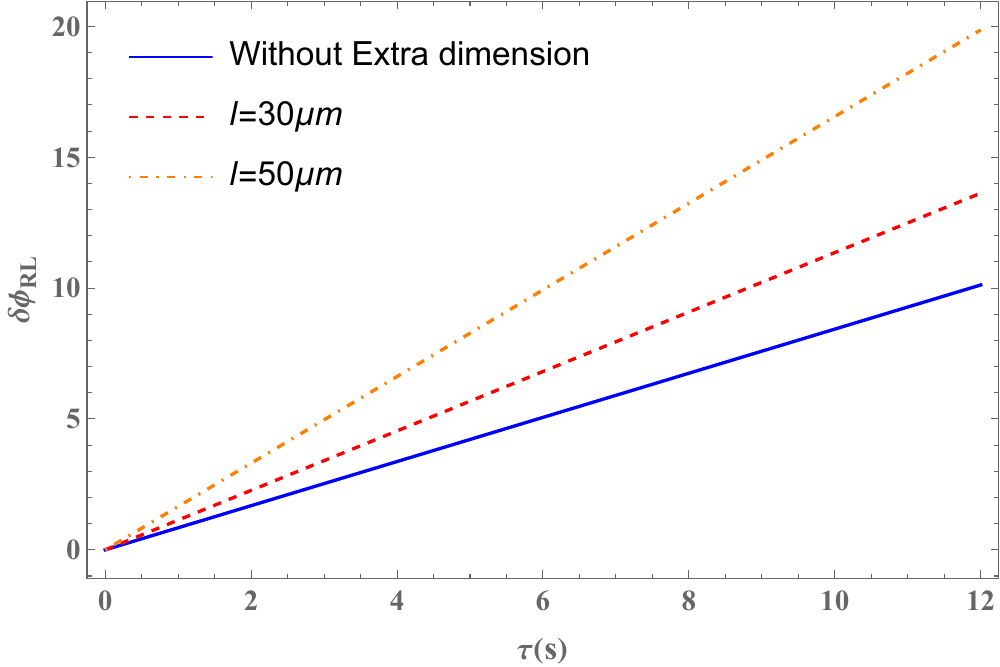}
		\caption{}
		\label{fig:b}
	\end{subfigure}
	\begin{subfigure}[t]{0.4\textwidth}
		\centering
		\includegraphics[width=7.0cm]{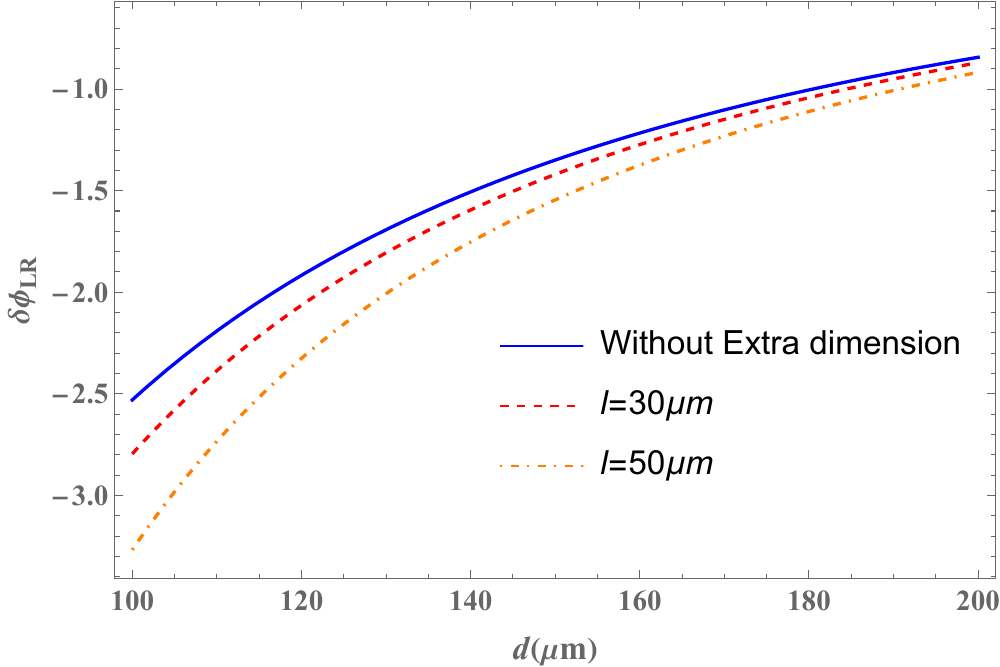}
		\caption{}
		\label{fig:c}
	\end{subfigure}
	\begin{subfigure}[t]{0.4\textwidth}
		\centering
		\includegraphics[width=7.0cm]{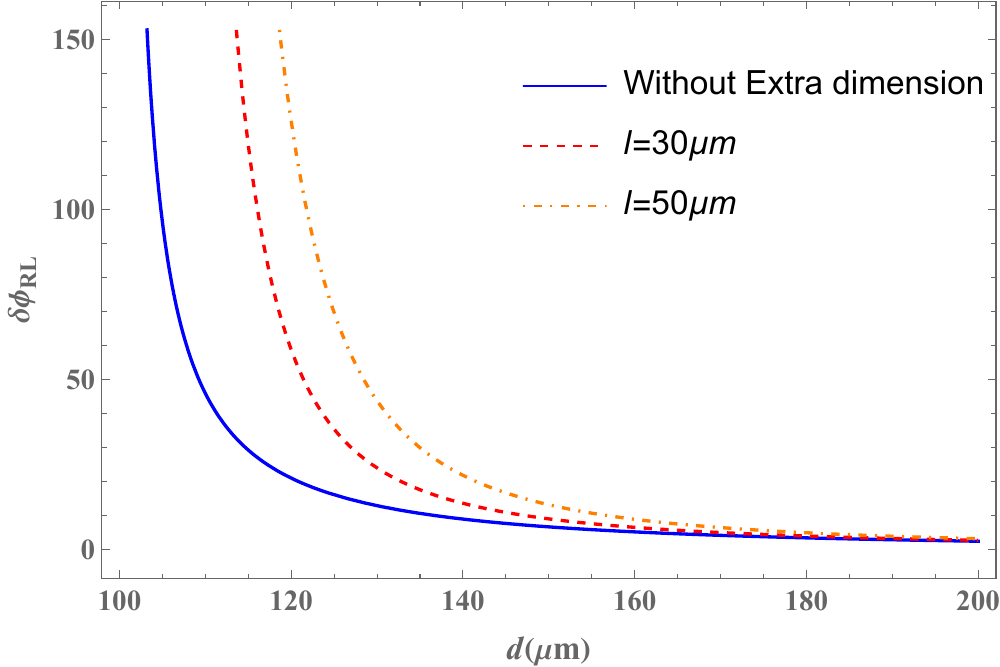}
		\caption{}
		\label{fig:d}
	\end{subfigure}
	\captionsetup{justification=RaggedRight}
	\caption{Comparison of the phase shifts with and without extra dimension, with $d = 150 \ \mathrm{\mu m}$ and $\delta x = 100 \ \mathrm{\mu m}$ for the upper two figures, and $t = 8 \ $s and $\delta x = 100 \ \mathrm{\mu m}$ for the lower two.}
	\label{fig2}
\end{figure*}

At starting time $\tau = 0$, the state is given by
\begin{equation}
	|\psi\left (\tau=0  \right )\rangle_{12}= \frac{1}{\sqrt{2}} \left (\ket{L }_{1}+\ket{R}_{1}\right) \frac{1}{\sqrt{2}}\left (\ket{L}_{2}+\ket{R }_{2} \right),
\end{equation}
where $\ket{L}$ and $\ket{R}$ represent spatially localized states of two masses in superposition.  After passing through the device, spin-up states will appear on the left, while spin-down states will appear on the right. So after time t, we get
by \cite{boseSpinEntanglementWitness2017}
\begin{equation}
	\begin{split}		
	|\psi\left (\tau= t  \right )\rangle_{12} =& \frac{1}{2} \Big [\ket{L,\uparrow }_{1}\left (\ket{L,\uparrow}_{2}+e^{i\delta \Phi_{LR}}\ket{R,\downarrow}_{2} \right )  \\ +
	&  \ket{R,\downarrow}_{1}\left (e^{i\delta \Phi_{RL}}\ket{L,\uparrow}_{2}+\ket{R,\downarrow }_{2} \right )  \Big ]\ket{C}_{1}\ket{C}_{2},
\end{split}  
\end{equation}
Therefore, only spin will be used to represent the state of the particle. The time evolution of the entanglement state of the two masses at $ \tau=\tau_{end}$ is given by \cite{boseSpinEntanglementWitness2017}
\begin{equation}	
	\begin{split}		
		|\psi\left (\tau=\tau_{end}  \right )\rangle_{12} =& \frac{1}{2} \Big [\ket{\uparrow }_{1}\left (\ket{\uparrow}_{2}+e^{i\delta \Phi_{LR}}\ket{\downarrow}_{2} \right )  \\ +
		&  \ket{\downarrow}_{1}\left (e^{i\delta \Phi_{RL}}\ket{\uparrow}_{2}+\ket{\downarrow }_{2} \right )  \Big ]\ket{C}_{1}\ket{C}_{2},
	\end{split}  		
	\label{states}
\end{equation}

where $|C\rangle_{j}$ represents the initial motional states of masses, the phase of a quantum state is determined by $e^{-i(E/\hbar)t}$, which can be expressed as $\phi=(E/\hbar)t$. In the case of gravitational interaction alone, the energy is given by $E=Gm_{1}m_{2}/d$, resulting in a phase shift $\Phi=\frac{Gm_{1}m_{2}t}{\hbar d}$ \cite{christodoulou2019possibility}.
Without an extra dimension, the phase shifts can be computed as follows:
\begin{eqnarray}
	\delta \Phi _{LR}= \Phi _{LR}- \Phi,\quad \delta \Phi _{RL}= \Phi _{RL}- \Phi,
\end{eqnarray}
where the phases are given by:
\begin{equation}
	\begin{aligned}
		&\Phi \sim \frac{Gm_{1}m_{2}\tau }{\hbar d}, \qquad
		\Phi _{RL}\sim \frac{Gm_{1}m_{2}\tau }{\hbar\left ( d-\delta x \right )},\\
		&\Phi _{LR}\sim \frac{Gm_{1}m_{2}\tau }{\hbar\left ( d+\delta x \right )},
	\end{aligned}
\end{equation}
where $d$ represents the distance between the two masses, and $\delta x$ represents the interval between $\ket{L}$ and $\ket{R}$.

To investigate the impact of extra dimension, we consider the corrected gravitational potential originated from extra dimension theory (\ref{correctGP}).
In such a case, the freely falling process is different from that in standard four-dimensional theory, and the entanglement state would contain the information of extra dimension.
we were able to ascertain the resulting phase shift with extra dimension \cite {stodolsky1979matter}.

Specifically, they are contained in the phase shifts,
\begin{equation}
	\begin{aligned}
		&\tilde{\Phi} \sim \frac{Gm_{1}m_{2}\tau }{\hbar d}(1+\frac{2l^{2}}{3d^{2}}),\\
		&\tilde{\Phi}_{RL}\sim \frac{Gm_{1}m_{2}\tau }{\hbar\left ( d-\delta x \right )}(1+\frac{2l^{2}}{3\left ( d-\delta x \right )^{2}}),\\
		&\tilde{\Phi}_{LR}\sim \frac{Gm_{1}m_{2}\tau }{\hbar\left ( d+\delta x \right )}(1+\frac{2l^{2}}{3\left ( d+\delta x \right )^{2}}).
	\end{aligned}
	\label{shifts}
\end{equation}
Furthermore, the phase shifts are given by:
\begin{eqnarray}
	&&\delta \tilde{\Phi}_{RL}= \tilde{\Phi}_{RL}- \tilde{\Phi},\label{phaseshift1}
	\\
	&&\delta \tilde{\Phi}_{LR}= \tilde{\Phi}_{LR}- \tilde{\Phi}.\label{phaseshift2}
\end{eqnarray}

In our analysis, we focus on microobjects with a mass on the order of $10^{-14}$ kg. To ensure that the AdS radius, denoted as $l$, is below the current lower limit for testing the Newton's inverse square law, we set $l$ to be sufficiently small. The corresponding phase shifts are illustrated in Figure \ref{fig2}.



We observe that the phases change more rapidly in the existence of an extra dimension, indicating an accelerated evolution of the quantum states. As the radius of curvature increases, the magnitude of the phase changes becomes greater. This is due to the fact that the correction to the gravitational potential is proportional to the AdS radius $l$. Furthermore, the phase shift $\delta \Phi_{RL}$ undergoes more significant changes compared to $\delta \Phi_{LR}$. This is because the distance between the states $\ket{R}$ and $\ket{L}$ is relatively shorter, resulting in a weaker gravitational interaction and larger phase changes. Overall, these observations suggest that the presence of an extra dimension has significant effects on the phase shifts.

\begin{figure}[h!]
	\centering
	\includegraphics[width=8cm]{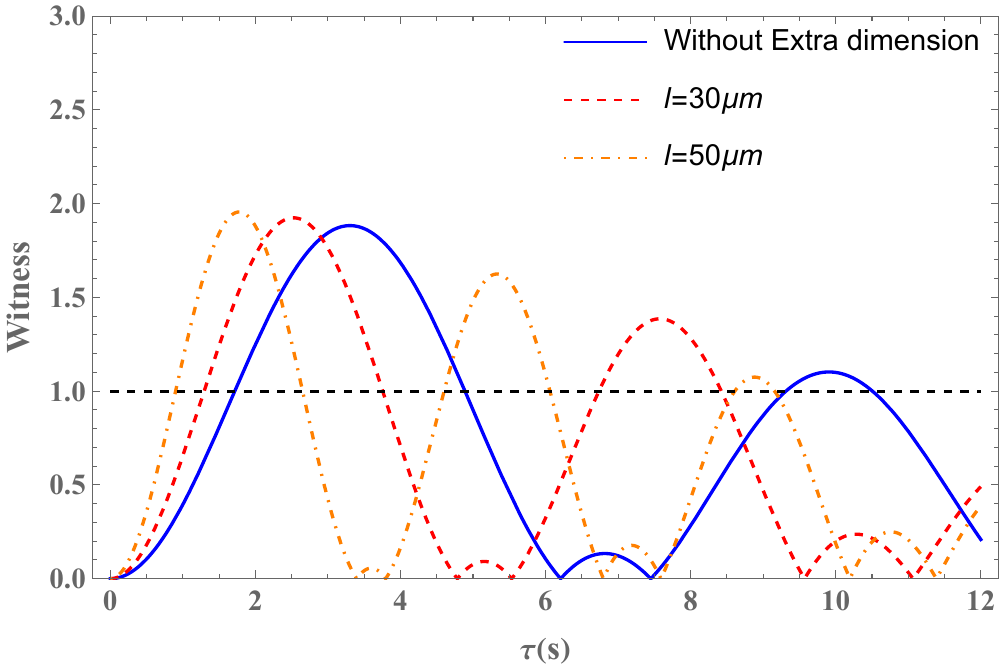}
	\includegraphics[width=8cm]{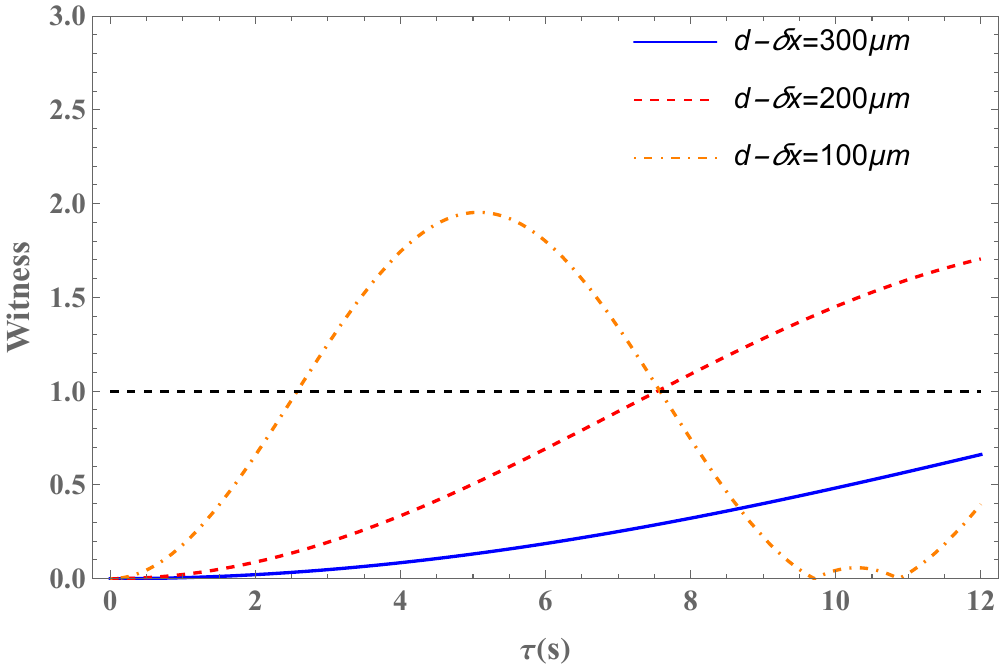}
	\includegraphics[width=8cm]{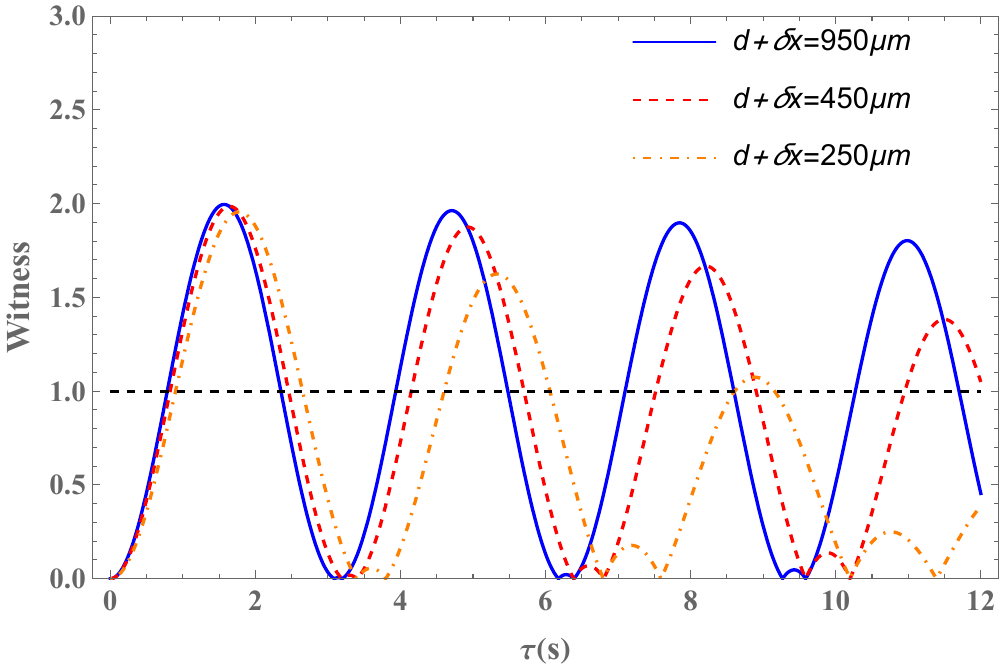}
	\captionsetup{justification=RaggedRight}
	\caption{We compare the entanglement witness $\mathscr{W}$ in the presence and absence of an extra dimension. In the top figure, the parameters used are $d=150~\mathrm{\mu m}$ and $\delta x=100~\mathrm{\mu m}$. For the middle figure, we set $d+\delta x=700~\mathrm{\mu m}$ and $l=50~\mathrm{\mu m}$. In the bottom figure, the values chosen are $d-\delta x=50~\mathrm{\mu m}$ and $l=50~\mathrm{\mu m}$. Instead of using just $d$ and $\delta x$, we employ $d+\delta x$ and $d-\delta x$ because they represent the farthest and closest distances between the two masses, respectively, as illustrated in Fig.~\ref{fig1}.}
	\label{fig3}
\end{figure}

To determine whether the masses are in entangled states, we need to calculate the witness generated by the two masses. The witness is defined as \cite{boseSpinEntanglementWitness2017}:
\begin{equation}
	\mathscr{W}=\left | \langle \sigma_{x}^{(1)}\otimes \sigma_{z}^{(2)} \rangle +\langle \sigma_{y}^{(1)}\otimes \sigma_{y}^{(2)} \rangle \right |,
\end{equation}
where $\sigma_{x,y,z}^{i}$ represents the Pauli matrix and
\begin{equation}
	\langle \sigma_{x}^{(1)} \otimes \sigma_{z}^{(2)} \rangle= \bra{\psi}\sigma_{x}^{(1)} \otimes \sigma_{z}^{(2)}\ket{\psi}.
\end{equation}
The phase shifts are incorporated through equation (\ref{states}). If $\mathscr{W}>1$, it indicates that the masses are in entangled states. We illustrate the witness $\mathscr{W}$ for various parameter choices using the obtained phase shifts in Fig.~\ref{fig3}.   Due to the current experimental observations \cite{lee2020new}, we chose $l \le  50 \mu \text{m}$.

In the upper part of Fig.~\ref{fig3}, we can observe that the witness of the masses reaches 1 more rapidly in the presence of an extra dimension. Notably, this behavior shows a positive correlation with the AdS radius. This correlation can be understood by examining the gravitational potential (\ref{correctGP}) and the phase shifts (\ref{shifts}), which both include correction terms proportional to $l^2$.

%
%

The middle figure with a fixed $d+\delta x$ and the bottom figure with a fixed $d-\delta x$ illustrate the effects of $d-\delta x$ and $d+\delta x$, respectively. From the middle figure, it is evident that the witness approaches unity rapidly for small values of $d-\delta x$. This behavior can be explained by examining the expressions (\ref{shifts}), (\ref{phaseshift1}), and (\ref{phaseshift2}), where $d-\delta x$ appears in the denominators. Consequently, the witness is sensitive to variations in $d-\delta x$.

In contrast, the bottom figure shows that the witness approaches unity more rapidly for larger values of $d+\delta x$, contrary to the behavior observed in the middle figure. This behavior also stems from equations (\ref{shifts}), (\ref{phaseshift1}), and (\ref{phaseshift2}). Although the distance $d+\delta x$ appears in the denominator, it should be noted that $\tilde{\Phi}_{LR}- \tilde{\Phi}<0$ according to equation \ref{phaseshift2}. Therefore, as the distance $d+\delta x$ increases, the entanglement phenomenon between the particles becomes more pronounced. This may be because the increase in $d+\delta x$ reduces the impact on the nearest pair, as depicted in Fig.~\ref{fig:c}.

When comparing the middle and bottom figures, we observe that the distance $d-\delta x$ has a more significant impact on the witness compared to $d+\delta x$. This is due to the fact that the interaction between $\ket{R}_{1}$ and $\ket{L}_{2}$ occurs over a shorter distance, $d-\delta x$. As a result, $\tilde{\Phi}_{RL}$ dominates, as depicted in Fig.~\ref{fig1}.

\section{Conclusion}

In this paper, we focused on the QGEM experiment proposed by Bose et al. \cite{boseSpinEntanglementWitness2017}. Our main objective was to investigate the impact of an extra dimension on the QGEM experiment. The presence of the extra dimension leads to corrections in the gravitational potential, which, in turn, affects the quantum phase and entanglement phenomena of the test masses.

By analyzing the model parameters, we gained insights into the effect of the extra dimension on particle entanglement. Our results demonstrate that in the presence of an RS-II type of extra dimension, compared to the four-dimensional theory,  the particle's witness will reach 1 in a shorter time. This acceleration can be attributed to the strengthening of the gravitational potential on short scales due to the presence of the extra dimension. It is important to note that these conclusions are specific to the RS-II braneworld model and may not necessarily hold true for other models.

Lastly, we acknowledge that our theoretical analysis in this paper neglected certain details of the experiment, such as the specific setup of distances and the Casimir-Polder force. Nevertheless, this work also presents a potential approach for detecting extra dimensions.

\section*{Acknowledgements}
This work is supported in part by the National Natural Science Foundation of China (Grants No.  11975116, 12165013). B.-M. Gu also acknowledges the support of Natural Science Foundation of Jiangxi province, China, under Grant No. 20224BAB211026.


apsrev4-2.bst 2019-01-14 (MD) hand-edited version of apsrev4-1.bst
Control: key (0)
Control: author (8) initials jnrlst
Control: editor formatted (1) identically to author
Control: production of article title (0) allowed
Control: page (0) single
Control: year (1) truncated
Control: production of eprint (0) enabled

\end{document}